\documentclass[12pt]{article}

\usepackage[margin=1in]{geometry}  	
\usepackage{amsfonts}
\usepackage{amsmath} 
\usepackage{booktabs}
\usepackage{multirow}
\usepackage[doublespacing]{setspace} 
\usepackage{float}
\usepackage{caption}
\usepackage{graphicx}
\usepackage{adjustbox}

\usepackage{lscape}
\usepackage{afterpage}

\usepackage{adjustbox}

\usepackage{xcolor}

\usepackage{relsize}

\usepackage{bbm}
\usepackage{subfig}

\usepackage{tikz}
\usetikzlibrary{positioning,shapes.geometric}

\usepackage{setspace}
\usepackage{cite}

\usepackage{rotating}

\definecolor{forestgreen}{RGB}{34,139,34}
\usepackage{hyperref}
\hypersetup{
    colorlinks=true,
    linkcolor=magenta,
    filecolor=magenta, 
    citecolor=forestgreen,      
    urlcolor=blue
}

\usepackage{amsmath,amsfonts,amssymb,amsthm}

\usepackage{authblk}

\DeclareMathOperator{\E}{\mbox{E}}

\usepackage[utf8]{inputenc}
\DeclareUnicodeCharacter{200E}{}

\def \sizevar {0.7}

\usepackage{datetime}

\usepackage[absolute,showboxes]{textpos}

\TPMargin{5pt}

\newcommand{\copyrightstatement}{
    \begin{textblock}{0.84}(0.08,0.93)    
         \noindent
         \footnotesize
         This draft manuscript presents work in progress. \\
         Comments and reports of mistakes are very much welcome at \href{mailto:issa\_dahabreh@brown.edu}{issa\_dahabreh@brown.edu}.
    \end{textblock}
}

\def\paperversionmajor{29}
\def\paperversionminor{0}

\usepackage{enumitem}%

\makeatletter
\def\@seccntformat#1{\@ifundefined{#1@cntformat}%
   {\csname the#1\endcsname\quad}  
   {\csname #1@cntformat\endcsname}
}
\let\oldappendix\appendix 
\renewcommand\appendix{%
    \oldappendix
    \newcommand{\section@cntformat}{\appendixname~\thesection\quad}
}
\makeatother

\begin{document}

\title{Sensitivity analysis using bias functions for studies extending inferences from a randomized trial to a target population}

\author[1,2,3,4]{Issa J. Dahabreh}
\author[4,5]{James M. Robins}
\author[5]{Sebastien J-P.A. Haneuse}
\author[1,2]{Iman Saeed}
\author[1,2]{Sarah E. Robertson}
\author[6]{Elisabeth A. Stuart}
\author[4,5]{Miguel A. Hern\'an}

\affil[1]{Center for Evidence Synthesis in Health, Brown University School of Public Health, Providence, RI}
\affil[2]{Department of Health Services, Policy \& Practice, School of Public Health, Brown University, Providence, RI}
\affil[3]{Department of Epidemiology, School of Public Health, Brown University, Providence, RI}
\affil[4]{Department of Epidemiology, Harvard T.H. Chan School of Public Health, Harvard University, Boston, MA}
\affil[5]{Department of Biostatistics, Harvard T.H. Chan School of Public Health, Harvard University, Boston, MA}
\affil[6]{Department of Pharmacy Practice, College of Pharmacy, University of Rhode Island, RI}
\affil[6]{Departments of Mental Health, Biostatistics, and Health Policy and Management, Johns Hopkins Bloomberg School of Public Health, Baltimore, MD}

\copyrightstatement

\maketitle{}
\thispagestyle{empty}

\clearpage
\pagenumbering{gobble}

\vspace*{1in}

\begin{abstract}
\noindent
\linespread{2}\selectfont
Extending (generalizing or transporting) causal inferences from a randomized trial to a target population requires ``generalizability'' or ``transportability'' assumptions, which state that randomized and non-randomized individuals are exchangeable conditional on baseline covariates. These assumptions are made on the basis of background knowledge, which is often uncertain or controversial, and need to be subjected to sensitivity analysis. We present simple methods for sensitivity analyses that do not require detailed background knowledge about specific unknown or unmeasured determinants of the outcome or modifiers of the treatment effect. Instead, our methods directly parameterize violations of the assumptions using bias functions. We show how the methods can be applied to non-nested trial designs, where the trial data are combined with a separately obtained sample of non-randomized individuals, as well as to nested trial designs, where a clinical trial is embedded within a cohort sampled from the target population. We illustrate the methods using data from a clinical trial comparing treatments for chronic hepatitis C infection.
\end{abstract}

\clearpage
\pagenumbering{arabic}

The distribution of effect modifiers among randomized individuals in clinical trials is often different from that of individuals seen in clinical practice; consequently, average treatment effects estimated in clinical trials do not directly apply to target populations beyond the population represented by the randomized individuals \cite{Rothwell2005}. Methods for extending (``generalizing'' or ``transporting'' \cite{hernan2016discussionKeiding}) causal inferences from a randomized trial to a target population require exchangeability assumptions (``generalizability'' or ``exchangeability'' assumptions), which state that randomized and non-randomized groups are exchangeable conditional on baseline covariates \cite{Cole2010, Westreich2017, dahabreh2018generalizing, dahabreh2018transporting}. These assumptions are made on the basis of background knowledge, which is often uncertain or controversial. Thus, investigators interested in extending inferences beyond the population of randomized individuals need to conduct sensitivity analyses to examine how potential violations of the assumptions would affect their findings.

The literature on sensitivity analysis for unmeasured confounding in observational studies or for the related problem of data missing not-at-random is very extensive (starting with \cite{cornfield1959smoking} and expanded in various ways, e.g., \cite{rosenbaum1983assessing,robins2000c, lash2011applying}). In the context of analyses extending inferences from a randomized trial to a target population, however, the only proposal for sensitivity analysis methods that we are aware of is the work of Nguyen et al. \cite{nguyen2017sensitivity, nguyen2018sensitivity}. Their approach can be useful when background knowledge is strong enough to suggest that a single variable that would render randomized and non-randomized groups exchangeable has been measured among the former but not among the latter. Their approach is less useful, however, in the more typical case where violations of the exchangeability assumptions are due to high dimensional unknown or unmeasured variables.

In this paper, we propose methods for sensitivity analysis that do not require detailed background knowledge about or measurement of \emph{specific} effect modifiers or determinants of the outcome. Instead, our methods parameterize violations of the exchangeability assumption using bias functions expressed in terms of differences between the potential (counterfactual) outcome means of randomized and non-randomized individuals. 

To illustrate the methods, we take advantage of data from two centers participating in the HALT-C trial comparing peginterferon alfa-2a treatment versus no treatment for patients with chronic hepatitis C infection. We treat one of the centers as the ``index center'' and the other as the ``target center,'' the latter providing a sample of individuals from a population different from the index center (see \cite{rudolph2017} for a similar setup). Our goal is to transport causal inferences from the index center to the population represented by the target center under a mean exchangeability assumption, and to propose methods for sensitivity analysis when the assumption does not hold. Because both centers were actually participating in the same trial, and treatment and outcome data are available from both, we have the ability to compare the estimates from our analyses against the randomization-based analyses in the target center. The data structure in our application is the same as in any \emph{non-nested trial design}, where a clinical trial dataset is combined with a separately obtained sample from a population of non-randomized individuals \cite{dahabreh2019studydesigns}. In the Appendix, we show how our methods can be extended to \emph{nested trial designs}, in which the randomized trial is nested within a cohort sampled from the target population \cite{dahabreh2018generalizing, dahabreh2019studydesigns}.

\section{Extending inferences to a target population}\label{section:identifiability}

Consider a \emph{non-nested trial design}, consisting of a randomized trial and a separately obtained simple random sample of non-randomized individuals from the target population \cite{dahabreh2018transporting, dahabreh2019studydesigns}. The data from this design consist of independent observations on baseline covariates, $X$; time-fixed (non-time-varying) treatments, $A$; the outcome, $Y$; and a trial participation indicator, $S$ (1 for randomized or 0 for non-randomized individuals). The data exhibit a special missingness pattern: for randomized individuals we have data on $(S =1 , X, A, Y);$ for non-randomized individuals we only have data on $(S = 0, X)$ \cite{dahabreh2018generalizing, dahabreh2018transporting}.

Let $Y^a$ be the potential (counterfactual) outcome under intervention to set treatment to $a$ \cite{splawaneyman1990, rubin1974}. We only consider binary treatments, such that $ a \in \{0,1\}$; extensions to multivalued treatments are straightforward. In the main text of this paper, the causal contrast of interest is the average treatment effect in the subset of non-randomized individuals in the target population, $ \E [Y^1 - Y^0 | S =0 ],$ which is identifiable both in non-nested and nested trial designs. In general, this treatment effect is not equal to the treatment effect among randomized individuals, $ \E [Y^1 - Y^0 | S = 0] \neq \E [Y^1 - Y^0 | S =1] $. In nested trial designs, we can also identify the average treatment effect in the overall target population, $ \E [Y^1 - Y^0 ]$ \cite{dahabreh2018generalizing} (i.e., not just the non-randomized sub-population). In the main text of this paper, we focus on identification, estimation, and sensitivity analysis methods for $ \E [Y^1 - Y^0 | S = 0]$ because our motivating application only allows identification of that causal effect \cite{dahabreh2018transporting}; in the Appendix we propose sensitivity analysis methods for studies that aim to draw causal inferences about $ \E [Y^1 - Y^0 ]$ \cite{dahabreh2018generalizing}.

\subsection{Identifiability conditions}\label{section:identifiability}

We now discuss sufficient conditions for identifying the potential outcome means among non-randomized individuals, $\E [Y^a | S = 0]$. These means are of inherent scientific interest and can also be used to identify the causal contrast of interest, because  $ \E [Y^1 - Y^0 | S = 0] = \E [Y^1  | S = 0] - \E [Y^0  | S = 0]$.

\noindent
\emph{(I) Consistency:} The observed outcome for the $i$th individual who received treatment $a$ equals that individual's counterfactual outcome under the same treatment, that is, $ \mbox{if } A_i = a, \mbox{ then } Y_i = Y^{a}_i.$ Implicit in this notation is the assumption that the invitation to participate in the trial and trial participation do not affect the outcome except through treatment assignment. In effect, we are making an exclusion restriction assumption of no direct effect of trial participation on the outcome, such that potential outcomes need only be indexed by treatment $a$, not trial participation $s$.

\noindent
\emph{(II) Mean exchangeability in the trial (over $A$):} Among randomized individuals, the potential outcome mean under treatment $a$ is independent of treatment, conditional on baseline covariates, $ \E [ Y^{a} | X, S = 1, A = a] = \E [ Y^{a} | X , S = 1 ].$

\noindent
\emph{(III) Positivity of treatment assignment:} In the trial, the probability of being assigned to each treatment, conditional on the covariates needed for exchangeability, is positive: $ \Pr[A=a | X = x, S=1] > 0 $ for every $a$ and every $x$ with positive density among randomized individuals, $f_{X|S}(x| S = 1) > 0$.
 
\noindent
\emph{(IV) Mean transportability (exchangeability over S):} The potential outcome mean is independent of trial participation, conditional on baseline covariates, $ \E [ Y^{a} | X, S = 1 ] = \E [ Y^{a} | X, S = 0 ]$ (provided the conditional expectations are well-defined). For binary $S$, this assumption is equivalent to the mean generalizability assumption $ \E [ Y^{a} | X, S = 1 ] = \E [ Y^{a} | X]$.
 
\noindent
\emph{(V) Positivity of trial participation:} The probability of participating in the trial, conditional on the covariates needed to ensure conditional mean transportability, is positive, $\Pr[S=1 | X = x] >0$ for every $x$ with positive density in the population, $f_X(x) > 0.$

Note that we have used $X$ generically to denote baseline covariates. It is possible, however, that strict subsets of $X$ are adequate to satisfy each exchangeability condition. For example, in a marginally randomized trial, the mean exchangeability among trial participants holds unconditionally.

Consistency, mean exchangeability, and positivity of treatment assignment are expected to hold in (marginally or conditionally) randomized clinical trials of well-defined interventions. In order to focus our attention on issues related to selective trial participation, we assume complete adherence to treatment assignment and no loss to follow-up in the trial.

Mean transportability and positivity of trial participation are the assumptions that allow us to extend causal inferences beyond the randomized trial. Positivity of trial participation is, in principle, testable \cite{petersen2012diagnosing}. In contrast, the \emph{mean transportability condition is not testable using the observed data}; in many applications, it is a controversial or uncertain \emph{assumption}.

\subsection{Identification of potential outcome means}

The conditions listed above are sufficient to identify the conditional potential outcome mean in the population of non-randomized individuals using only the observed data \cite{dahabreh2018transporting},
\begin{equation}\label{eq:identitification_S0}
	\E [Y^a | S = 0] = \E\! \big[\! \E [Y | X , S = 1, A = a] \big| S = 0\big].
\end{equation}
In other words, provided that identifiability conditions I through V hold, the observed data functional, $\E \big[ \E [Y | X , S = 1, A = a ] \big| S = 0 \big],$ can be interpreted as the potential outcome mean had non-randomized individuals received treatment $a$. This functional can be re-expressed using inverse odds (IO) of participation weighting \cite{dahabreh2018transporting},
\begin{equation}\label{eq:ipw_identification_SO}
	\E\! \big[\! \E [Y | X , S = 1, A = a] \big | S = 0 \big] = \dfrac{1}{\Pr[S = 0]} \E  \left[ \dfrac{S I( A = a ) Y \Pr [ S = 0  | X ] }{ \Pr [ S =1  | X ] \Pr [A =a | X , S = 1 ]  } \right],
\end{equation}
where $I(A = a)$ is the indicator function that takes value 1 when $A = a$ and zero otherwise. Furthermore, under the positivity conditions, 
\begin{equation}\label{eq:ipw_Hajek_identification}
	\Pr[S = 0] = \E  \left[ \dfrac{S I( A = a ) \Pr [ S = 0  | X ] }{ \Pr [ S = 1  | X ] \Pr [A = a | X , S = 1 ]  } \right],
\end{equation}
which will be useful in deriving estimators for the functional. Note that in non-nested designs, expectations and probabilities are with respect to the distribution induced by the study design \cite{bickel1993efficient,breslow2000semi} (i.e., the separate sampling of randomized and non-randomized individuals, with unknown sampling probability for non-randomized individuals \cite{dahabreh2018transporting, dahabreh2019studydesigns}).

\subsection{Estimation and inference}\label{section:estimation}

\subsubsection{Estimation} 

We briefly review \cite{dahabreh2018transporting} estimators of potential outcome means and treatment effects in the population of non-randomized individuals. All estimators use data on $O_i$, $i = 1, \ldots, n$, where $n$ is the total number of randomized and non-randomized individuals contributing data to the analysis, and
\begin{equation*}
  O =   \begin{cases}
      (X, S = 1, A, Y) \mbox{, for randomized individuals;} \\
      (X, S =0) \mbox{, for non-randomized individuals.}
    \end{cases}
\end{equation*}

\textbf{Potential outcome means:} We can estimate potential outcome means using estimators that rely on modeling the outcome mean, the probability of trial participation, or both. 

\noindent
\emph{Outcome model-based (g-formula) estimator:} We can use the sample analog estimator based on (\ref{eq:identitification_S0}): 
\begin{equation}\label{estimator:outcome_model}
	\widehat \mu_{\text{\scalebox{\sizevar}{OR}}}(a) = \Bigg\{\sum\limits_{i=1}^{n}(1 - S_i)\Bigg\}^{-1} \sum\limits_{i=1}^{n} (1 - S_i) \widehat g_a(X_i),
\end{equation}
where $\widehat g_a(X)$ is an estimator for $\E[Y| X, S = 1, A = a]$. In applied analyses, it is impossible to nonparametrically estimate this conditional mean because of the curse of dimensionality \cite{robins1997toward}, and we need to make modeling assumptions. Typically, we posit a parametric model $g_a(X; \theta)$, with finite dimensional parameter $\theta$. When using such a model, the validity of the g-formula estimator depends on correct model specification, in addition to the identifiability conditions.

\noindent
\emph{IO weighting:} Using (\ref{eq:ipw_identification_SO}), we can obtain an IO weighting estimator,
\begin{equation} \label{estimator:iow1}
	\widehat \mu_{\text{\scalebox{\sizevar}{IOW1}}}(a) = \Bigg\{\sum\limits_{i=1}^{n}(1 - S_i)\Bigg\}^{-1} \sum\limits_{i=1}^{n} \widehat w_a(X_i, S_i, A_i)  Y_i ,
\end{equation}
where $$\widehat w_a(X_i, S_i, A_i) = S_i I(A_i = a) \dfrac{1 - \widehat p(X_i)  }{\widehat p(X_i) \widehat e_a(X_i)},$$ $\widehat p(X)$ is an estimator for $\Pr[S = 1 | X]$, and $\widehat e_a(X)$ is an estimator for $\Pr[A =a | X, S = 1]$.
Alternatively, combining (\ref{eq:ipw_identification_SO}) and (\ref{eq:ipw_Hajek_identification}), we can normalize the weights to sum to 1,
\begin{equation} \label{estimator:iow2}
	\widehat \mu_{\text{\scalebox{\sizevar}{IOW2}}}(a) = \left\{ \sum\limits_{i=1}^{n} \widehat w_a(X_i, S_i, A_i)  \right\}^{-1} \sum\limits_{i=1}^{n} \widehat w_a(X_i, S_i, A_i) Y_i .
\end{equation}
As for the outcome model-based estimator, in applications, it is impossible to nonparametrically estimate the conditional probability of trial participation $\Pr[S = 1 | X]$ and we have to make modeling assumptions. Typically, we posit a parametric model $p(X; \beta)$ for $\Pr[S = 1 | X]$, with finite-dimensional parameter $\beta$. When using such a model, the validity of the IO weighting estimators depends on correct model specification, in addition to the identifiability conditions. The conditional probability of treatment among randomized individuals $\Pr[A = a | X, S = 1]$ is known and does not have to be estimated; in the presence of baseline covariate imbalances between the randomized groups, however, it is useful to estimate it using a parametric model, say, $e_a(X; \gamma)$, with finite-dimensional parameter $\gamma$ \cite{robins1994estimation, robins1995semiparametric, hahn1998role, lunceford2004, williamson2014variance}.

\noindent
\emph{Augmented IO weighting:} To improve the efficiency of the IO weighting estimator and gain robustness to misspecification of the models for the conditional outcome mean or the probability of trial participation, we can use the augmented (doubly robust) IO weighting estimator
\begin{equation} \label{estimator:aiow1}
\begin{split}
\widehat{\mu}_{\text{\scalebox{\sizevar}{AIOW1}}} (a)	 =  \Bigg\{\sum\limits_{i=1}^{n}(1 - S_i)\Bigg\}^{-1}  \sum_{i=1}^{n} \left\{\widehat w_a(X_i, S_i, A_i) \big\{ Y_{i}   -   \widehat g_a (X_i)  \big\} +  (1 - S_i) \widehat g_a (X_i)  \right\}.
\end{split}
\end{equation}
Again, we can normalize the weights to sum to 1,
\begin{equation} \label{estimator:aiow2}
\begin{split}
\widehat{\mu}_{\text{\scalebox{\sizevar}{AIOW2}}} (a) &=  \left\{ \sum_{i = 1}^{n} \widehat w_a(X_i, S_i, A_i) \right\}^{-1}  \sum_{i=1}^{n}    \widehat w_a(X_i, S_i, A_i)  \big\{ Y_{i} - \widehat g_a (X_i) \big\}  \\
 & \quad\quad\quad\quad\quad\quad+  \Bigg\{\sum\limits_{i=1}^{n}(1 - S_i)\Bigg\}^{-1}   \sum_{i = 1}^{n} (1 - S_i) \widehat g_a (X_i).
\end{split}
\end{equation}

These two estimators are ``doubly robust'' in the sense that they produce valid results when either the model for the probability of trial participation or the model for the conditional outcome mean among randomized individuals is correctly specified.

\textbf{Treatment effects:} We can use the potential outcome mean estimators to estimate the average treatment effect in the target population of non-randomized individuals, $\E [Y^1 - Y^0 | S = 0]$, by differencing. For instance, we can estimate the average treatment effect using the augmented IO weighting estimator with normalized weights as $\widehat \mu_{\text{\scalebox{\sizevar}{AIOW2}}}(1) - \widehat \mu_{\text{\scalebox{\sizevar}{AIOW2}}}(0).$

\subsubsection{Inference} 

Confidence intervals for the estimated potential outcome means and mean differences can be obtained by the usual ``sandwich'' approach \cite{Stefanski2002} or by bootstrap methods \cite{efron1994introduction}. Both approaches can account for uncertainty in estimating the parameters of the model for the probability of trial participation or the model for the outcome mean.

\section{Sensitivity analysis for violations of the transportability assumption}

\subsection{Violations of the transportability assumption}

The validity of all the estimators in the previous section depends critically on the mean transportability assumption. In most applications, however, it is unlikely that we know or can measure enough baseline variables to ensure that the assumption holds. Thus, we need to consider the impact of assumption violations, when
\begin{equation*}
	\E[Y^a | X , S = 1 ] \neq \E[Y^a | X , S = 0 ].
\end{equation*}
The magnitude of the violations, that is, the magnitude of the difference between the conditional potential outcome means $\E[Y^a | X , S = 1 ]$ and $\E[Y^a | X , S = 0 ]$, determines the amount of bias. Because this magnitude cannot be assessed using the data, we need to conduct sensitivity analyses to examine the impact of violations of the condition affect on our results \cite{robins2000c}.

\subsection{Sensitivity analysis with bias functions}

\subsubsection{Bias functions}

Following prior work on sensitivity analysis for marginal structural causal models \cite{robins2000c,brumback2004sensitivity,robins1999association}, we parameterize violations of the mean transportability condition using a ``bias function'' for each treatment $a$, defined as 
\begin{equation} \label{eq:parameterization}
	u(a, X ) 	\equiv \E[Y^a | X , S = 1 ] - \E[Y^a | X , S = 0 ].
\end{equation}
Intuitively, $u(a, X)$ expresses violations of the transportability assumption for each treatment $a$ as a function of the potential outcome means between randomized and non-randomized individuals, conditional on (within strata of) the measured covariates $X$.

Parameterizing the violations of the transportability assumption allows us to re-express the conditional potential outcome mean under treatment $a$ among non-randomized individuals. First, by re-arranging the definition in (\ref{eq:parameterization}), we obtain $\E[Y^a | X, S = 0] = \E[Y^a | X, S = 1] - u(a, X).$ Next, under consistency (condition I), exchangeability of the treated and untreated groups in the trial (condition II), and positivity of treatment assignment in the trial (condition III), we have that $\E[Y^a | X, S = 1] = \E [Y | X, S = 1 , A = a].$ Putting everything together,
\begin{equation}\label{eq:bias_correction_S0}
\E[Y^a | X, S = 0] = \E [Y | X, S = 1 , A = a] - u(a, X).
\end{equation}
Using the law of iterated expectation and (\ref{eq:bias_correction_S0}), 
\begin{equation}\label{eq:basis_for_sens_S0}
	\begin{split}
	\E [Y^a | S = 0 ] &= \E\! \big[\! \E [Y^a | X, S = 0 ] \big| S = 0 \big] \\
			&= \E\! \big[\! \E [Y | X, S = 1 , A = a] - u(a, X) \big| S = 0 \big] \\
			&= \E\! \big[\! \E [Y | X , S = 1, A = a] \big| S = 0 \big] - \E\! \big[ u(a, X ) | S = 0 \big],
	\end{split}
\end{equation}
where the first term in the last expression is identifiable from the data, \emph{regardless of whether the transportability condition holds}, and the second term is also identifiable for each user-specified $u(a, X)$ function.

\subsubsection{Sensitivity analysis}\label{subsec:estimation_sensitivity}

The result in (\ref{eq:basis_for_sens_S0}) suggests simple approaches for sensitivity analysis using the potential outcome mean estimators in the previous section.

\noindent
\emph{Outcome model-based (g-formula) estimator:} We can modify the outcome model-based estimator to directly incorporate the bias correction,
\begin{equation}\label{estimator:or_bc_S0}
	\widehat \mu_{\text{\scalebox{\sizevar}{OR}}}^{\text{\scalebox{\sizevar}{bc}}}(a) = \left\{ \sum\limits_{i=1}^{n} (1 - S_i)  \right\}^{-1}  \sum\limits_{i=1}^{n}  (1 - S_i) \big\{ \widehat g_a(X_i) - u(a, X_i) \big\} .
\end{equation}

\noindent
\emph{IO weighting:} Similarly, we can modify the IO weighting estimators as
\begin{equation} \label{estimator:iow1_bc}
	\widehat \mu_{\text{\scalebox{\sizevar}{IOW1}}}^{\text{\scalebox{\sizevar}{bc}}}(a) = \left\{ \sum\limits_{i=1}^{n} (1 - S_i)  \right\}^{-1}  \sum\limits_{i=1}^{n}  \big\{\widehat w_a(X_i, S_i, A_i) Y_i  - (1 - S_i) u(a, X_i) \Big\} ,
\end{equation}
or, normalizing the weights to sum to 1, 
\begin{equation} \label{estimator:iow2_bc}
	\begin{split}
	\widehat \mu_{\text{\scalebox{\sizevar}{IOW2}}}^{\text{\scalebox{\sizevar}{bc}}}(a) &= \left\{ \sum\limits_{i=1}^{n} \widehat w_a(X_i, S_i, A_i) \right\}^{-1} \sum\limits_{i=1}^{n} \widehat w_a(X_i, S_i, A_i) Y_i   \\
		&\quad\quad\quad\quad\quad\quad\quad -  \left\{ \sum\limits_{i=1}^{n} (1 - S_i)  \right\}^{-1}  \sum\limits_{i=1}^{n}  (1 - S_i) u(a, X_i).
		\end{split}
\end{equation}
In the Appendix, we show how the bias correction for the IO weighting estimators can be easily implemented using standard regression software, by simply re-coding the outcome values to directly incorporate the bias correction function $u(a, X)$.

The validity of the estimators in equations (\ref{estimator:or_bc_S0}) through (\ref{estimator:iow2_bc}) depends on the correct choice of the bias functions $u(a, X)$ and, when relying on parametric models, the correct specification of the models for the mean outcome among randomized individuals, $g_a(X; \theta)$, or the probability of trial participation conditional on covariates, $p(X; \beta)$. By \emph{correct specification} we mean that the models $g_a(X; \theta)$ and $p(X; \beta)$ should be good approximations of the corresponding conditional mean/probability functions. Correct model specification is distinct from the mean transportability condition: informally, correct model specification addresses differences between randomized and non-randomized individuals with respect to \emph{measured variables}; the bias functions address residual differences due to \emph{unmeasured or unknown variables}.

\noindent
\emph{Augmented IO weighting:} We can incorporate the bias correction functions in the augmented IO weighting estimators,
\begin{equation} \label{estimator:aiow11}
\begin{split}
\widehat{\mu}_{\text{\scalebox{\sizevar}{AIOW1}}}^{\text{\scalebox{\sizevar}{bc}}} (a)	 &=  \left\{ \sum\limits_{i=1}^{n} (1 - S_i)  \right\}^{-1} \sum_{i=1}^{n} \Biggl\{ \widehat w_a(X_i, S_i, A_i)  \big\{ Y_i  - \widehat g_a (X_i)   \big\}  \\ 
&\quad\quad\quad\quad\quad\quad\quad\quad\quad\quad\quad\quad + (1 - S_i) \big\{ \widehat g_a (X_i) - u(a, X) \big\}  \Biggl\} ,
\end{split}
\end{equation}
or, normalizing the weights to sum to 1,
\begin{equation} \label{estimator:aiow2_bc}
\begin{split}
\widehat{\mu}_{\text{\scalebox{\sizevar}{AIOW2}}}^{\text{\scalebox{\sizevar}{bc}}}(a) &=  \left\{  \sum_{i = 1}^{n} \widehat w_a(X_i, S_i, A_i)  \right\}^{-1} \sum_{i=1}^{n}  \widehat w_a(X_i, S_i, A_i) \big\{ Y_{i} - \widehat g_a (X_i) \big\} \\
  &\quad\quad\quad\quad\quad\quad + \left\{ \sum\limits_{i=1}^{n} (1 - S_i)  \right\}^{-1} \sum\limits_{i=1}^{n} (1 - S_i) \big\{ \widehat g_a(X_i) - u(a, X_i)  \big\} .
\end{split}
\end{equation}

These estimators retain the double robustness property of their non-bias corrected counterparts, in the sense that, when the bias correction function is correctly specified, they produce valid results when either the model of the outcome mean among randomized individuals or the model for the probability of trial participation is correctly specified.

\textbf{Sensitivity analysis for treatment effects:} As when the transportability condition holds, we can perform sensitivity analysis for treatment effects by differencing the appropriate bias-corrected estimators.

\textbf{Inference for sensitivity analysis:} The general theory in \cite{robins2000c} shows that valid confidence intervals for the estimated bias-corrected potential outcome means and mean differences can also be obtained by ``sandwich'' or bootstrap \cite{efron1994introduction} methods, while accounting for the uncertainty in estimating the parameters of the model for the probability of trial participation.

In the Appendix we discuss the identification, estimation, and sensitivity analysis for potential outcome means and average treatment effects in the overall target population (i.e., not just the non-randomized subset of the population).

\subsection{Choosing bias functions in practice}

We cannot identify the bias functions $u(a, X)$ using the data. Instead, we can use different bias functions to conduct sensitivity analyses. To develop some intuition about the choice of functions, note that $u(a, X)$ quantifies the degree of selection into the trial on the basis of potential outcome $Y^a$. For instance, suppose that greater outcome values are preferred (e.g., the outcome is a quality of life score with higher values indicating higher quality). If we believe that, conditional on measured covariates, randomized individuals have better outcomes than non-randomized individuals in the absence of treatment, that is, we believe that $\E [Y^0 | X , S = 1] > \E [Y^0 | X , S = 0]$, then we should select $u(0, X) > 0$. Conversely, if we believe that randomized individuals have worse outcomes than non-randomized individuals in the absence of treatment, that is, we believe that $\E [Y^0 | X , S = 1] < \E [Y^0 | X , S = 0]$, then we should select $u(0, X) < 0$. 

Now, define the difference of the bias functions, $\delta(X)$,
\begin{equation*}
	\begin{split}
\delta(X) &\equiv u(1, X) - u(0, X) \\
	&= \Big\{ \E [Y^1 | X , S = 1] - \E [Y^1 | X , S = 0]  \Big\} - \Big\{ \E [Y^0 | X , S = 1] - \E [Y^0 | X , S = 0]  \Big\} \\
	&= \Big\{ \E [Y^1 | X , S = 1] - \E [Y^0 | X , S = 1]  \Big\} - \Big\{ \E [Y^1 | X , S = 0] - \E [Y^0 | X , S = 0]  \Big\} \\
	&= \E [Y^1 - Y^0| X , S = 1]  - \E [Y^1 - Y^0  | X , S = 0].
	\end{split}
\end{equation*}
This calculation shows that the difference of the bias functions equals the difference of the conditional average treatment effects given $X$ among trial participants and non-participants. In other words, the difference of the bias functions reflects the ``residual,'' unexplained by $X$, effect modification over the participation indicator. 

For instance, suppose, as above, that greater outcome values are preferred and that we believe that $\delta(X) > 0$, that is, $u(1, X) > u(0, X)$. This corresponds to a belief that individuals who choose to participate in the trial benefit more (or are harmed less) from $a=1$ than $a=0$, compared to individuals who choose not to participate. A similar argument shows that, when higher values of the outcome are preferred, $\delta(X) < 0$ means that individuals who choose to participate in the trial benefit less (or be harmed more) from $a=1$ than $a=0$, compared to individuals who choose not to participate. In the special case of $\delta(X) = 0$, the conditional average treatment effect among randomized and non-randomized individuals is the same, meaning that trial participants are not selected based on the magnitude of the benefit (or harm) they might experience from treatment.

This interpretation of $u(0, X)$ and $\delta(X)$ suggests a convenient way to perform sensitivity analysis. For simplicity, we might use functions that do not depend on covariates, such that $u(0, X) \equiv u(0)$ and $\delta(X) \equiv \delta$. Clearly, each choice of a pair of values for $\big( u(0), \delta \big)$ implies a choice of $u(1) = \delta + u(0)$. The sensitivity analysis, then, examines the impact of a sufficiently diverse set of $\big( u(0), \delta \big)$ pairs on inferences about potential outcome means and average treatment effects, using the estimators in Section \ref{subsec:estimation_sensitivity}. 

Arguably, allowing the bias functions to vary over baseline covariates is more realistic. When supported by background knowledge, the use of the covariate-dependent functions $u(0, X)$ and $\delta(X)$, corresponding to a covariate-dependent $u(1, X)$, should result in more informative sensitivity analyses. When background knowledge is not that sharp, however, the simple approach of examining pairs of $\big( u(0), \delta \big)$ over a sufficiently broad range of values, may be adequate to  explore the impact of violations of the transportability condition on generalizability inferences. Of note, the double robustness property holds regardless of the dependence of the bias functions on covariates, provided equation (\ref{eq:parameterization}) is satisfied.

\section{Application to the HALT-C trial}

\subsection{Data and methods}

\paragraph{The HALT-C trial and transportability between centers:} The HALT-C trial enrolled patients with chronic hepatitis C and advanced fibrosis who had not responded to previous therapy and randomized them to treatment with peginterferon alpha-2a ($a=1$) versus no treatment ($a=0$). Patients were enrolled in 10 research centers and followed up every 3 months after randomization. We used data on the secondary outcome of platelet count at 9 months of follow-up; we report all outcome measurements as platelets $\times 10^3/$ml. To simplify exposition, we only used data from 205 patients (186 with complete data for our analyses) seen in two different research centers who had complete baseline covariate and outcome data: the \emph{index center}, $S = 1$, contributing 105 patients (94 with complete data); and the \emph{target center}, $S=0$, contributing 100 patients (92 with complete data). For purposes of illustration, we treated the target center data as a sample from a population of (trial-eligible) non-randomized individuals. Our goal was to transport causal inferences from the index center to the population represented by the target center. Because both centers were actually participating in the same trial, and because treatment and outcome data were available from both, we could compare estimates from transportability and sensitivity analyses against the randomization-based analyses in the target center. We purposely chose these two centers because they estimated the treatment effect on the platelet count to be substantially different (see below). In view of how we created the dataset for this illustration, our analyses should not be clinically interpreted.

\paragraph{Sensitivity analysis methods:} We used bias functions that where constant within levels of the baseline covariates. Specifically, we examined $u(0)$ values of $-40$, 0, and $+40$ and varied $\delta$ from $-60$ to $+60$, in steps of $20$, examining all possible $\big( u(0), \delta\big)$ pairs. We chose this range of values because, across all 10 centers participating in the HALT-C trial, the difference between the largest and smallest post-treatment mean was approximately $40$ for patients assigned to $a=1$ and 56 for patients assigned to $a=0.$ Furthermore, the standard deviation of the pre-treatment platelet count across all centers participating in the trial was approximately $66.1$. 

We used the bias functions with the estimators provided in the previous section to perform sensitivity analyses. We obtained confidence intervals for the potential outcome mean under each treatment using the robust (``sandwich'') variance \cite{Stefanski2002}, accounting for the estimation of the parameters of the working models. In Appendix \ref{appendix:covariate_dependent_bias_functions} we describe an additional example sensitivity analysis using covariate-dependent bias functions. We conducted all sensitivity analyses using the \texttt{R} \cite{rct2015} package \texttt{geex} \cite{saul2017}.

\paragraph{Model specification:} The sensitivity analysis methods require the specification of models for the outcome mean in each treatment group, the probability of trial participation, and the probability of treatment among randomized individuals. We specified logistic regression models for the probability of being in center $S=1$ and the probability of being assigned to peginterferon alpha-2a ($a=1$) among randomized individuals in that center. We also specified two linear regression models (one for each treatment arm) for the mean of the outcome among randomized individuals. Baseline covariate information is summarized in Table \ref{tab:descriptives}; all models used the baseline covariates listed in the table as main effects (baseline platelets, age, sex, treatment history, race/ethnicity, baseline white blood cell count, history of using needles or recreational drugs, transfusion history, body mass index, creatinine, and smoking). We built separate outcome models in each treatment group to allow for heterogeneity of the treatment effect over all baseline covariates included in the models.

\subsection{Results}

The ``base case'' analyses under mean transportability (i.e., $u(0) =0$ and $\delta=0$) produced similar results across different estimators (Table \ref{tab:treatment_effects_trans}), suggesting that models were approximately correctly specified \cite{Robins2001}. The unadjusted randomization-based analyses among patients with $S = 0$ produced fairly different results compared to the transportability analyses (Table \ref{tab:treatment_effects_no_trans}). The differences, however, were much smaller after using an augmented inverse probability weighting regression estimator to analyze the $S=0$ data (because of randomization among individuals in $S=0$, the covariate adjustment is virtually assumption-free \cite{tsiatis2008covariate}). The large change in estimates after covariate adjustment in the sample from $S=0$ is a reminder that large baseline covariate imbalances can occur in small samples despite randomization.

Sensitivity analysis results are summarized in Figure \ref{fig:sens_anal_HALT_C} and Appendix Figures \ref{fig:appendix_A1}. Overall, the results were only moderately sensitive to violations of the transportability assumption; for example, regardless of the choice of $u(0)$, values of $\delta$ smaller than -25 or -30 (depending on the estimator) would be needed for the estimated average treatment effect in $S=0$ to be in opposite direction compared to the base-case analysis (i.e., the black lines in the graphs cross only for $\delta$ values lower than -25 or -30). Sensitivity analysis results were much more uncertain when using IO weighting compared to the g-formula or augmented IO weighting estimators.

\section{Discussion}

We propose sensitivity analysis methods for violations of exchangeability assumptions in studies extending inferences from a randomized trial to the population of non-randomized individuals (in the main text) or the overall target population (in Appendix \ref{appendix:all_eligible}). The methods rely on specifying bias functions that directly parameterize violations of the required exchangeability assumptions. They can be applied to sensitivity analyses for outcome model-based (g-formula) estimators, probability of trial participation-based estimators (inverse probability or odds weighting), or augmented estimators that combine outcome and probability of participation models. Because of the additive structure of the bias correction, our methods are best-suited to continuous outcomes with unbounded support; a different sensitivity analysis method, suitable for non-continuous outcomes or continuous outcomes with bounded support, will be the subject of future work. 

The augmented weighting estimators are appealing for applied work because of their increased robustness to model misspecification. Because all methods depend critically on the specification of bias functions, which will often be highly speculative, some investigators might consider the double robustness property to be less compelling in the context of sensitivity analyses compared to when mean transportability holds. Even so, the augmented weighting estimators may be preferred compared to non-augmented weighting estimators because of improved efficiency; as illustrated in our analyses of data from the HALT-C trial. Furthermore, sensitivity analyses can be repeated with different bias functions to examine whether the choice of function leads to different conclusions regarding sensitivity to violations of the transportability assumption.

An attractive aspect of our methods is that they do not require detailed background knowledge about unknown or unmeasured variables. Instead, they only require expert judgments about the magnitude of the aggregate bias that these variables could induce. These judgments can be informed by examining readily available data on the variation of treatment effects among subgroups defined in terms of observed variables in the data at hand or external sources, including observational studies; the variation of treatment effects across studies examining similar interventions and outcomes in different populations (e.g., as assessed in meta-analyses); or the variation of the mean outcome under each treatments across populations and population subgroups. The benefit of our approach becomes clear when compared against approaches that require the specification of models for the distribution of unmeasured variables and the associations between unmeasured and measured variables (e.g., \cite{nguyen2017sensitivity, nguyen2018sensitivity}). These alternative approaches have multiple sensitivity parameters and require detailed background knowledge about sources of effect heterogeneity; such knowledge is often unavailable because empirical studies typically do not allow the precise assessment of effect modification \cite{dahabreh2016, kent2016risk}.

Some readers might find our sensitivity analysis approach unsatisfactory because it does not provide a single point estimate, and instead produces a range of results and associated confidence limits under possible violations of the transportability assumption \cite{robins2000c}. We believe that this is a desirable feature of our approach: when the transportability assumption does not hold, the data do not contain adequate information to identify the causal quantities of interest, and at best we can hope to examine how our conclusions would be affected by different violations of our assumptions. In a sense, we view sensitivity analysis as a way to encourage inferential modesty in analyses extending trial findings to a target population: sensitivity analysis highlights that the range of results compatible with the data, when considering violations of assumptions, is much broader than it appears when solely considering uncertainty due to sampling variability.

\section*{Acknowledgments}

The work was supported in part through Patient-Centered Outcomes Research Institute (PCORI) Methods Research Awards ME-1306­-03758 and ME­-1502­-27794 (PI: Dahabreh) and National Institutes of Health (NIH) grant R37 AI102634 (PI: Hern\'an). All statements in this paper, including its findings and conclusions, are solely those of the authors and do not necessarily represent the views of the PCORI, its Board of Governors, the PCORI Methodology Committee, or the NIH.
The illustrative data analyses in our paper used HALT-C research materials obtained from the NHLBI Biologic Specimen and Data Repository Information Coordinating Center and does not necessarily reflect the opinions or views of the HALT-C or the NHLBI.

\clearpage
\bibliographystyle{unsrt}
\bibliography{Transporting_the_results_of_experiments_sens_bias_cor}


\ddmmyyyydate 
\newtimeformat{24h60m60s}{\twodigit{\THEHOUR}.\twodigit{\THEMINUTE}.32}
\settimeformat{24h60m60s}
\begin{center}
\vspace{\fill}\ \newline
\textcolor{black}{{\tiny $ $transportability\_sens\_bias\_cor, $ $ }
{\tiny $ $Date: \today~~ \currenttime $ $ }
{\tiny $ $Revision: \paperversionmajor.\paperversionminor $ $ }}
\end{center}


\clearpage
\section*{Figure}

\begin{figure}[ht!]
	\centering
	\caption{Sensitivity analysis results for transporting inferences between two research centers participating in the HALT-C trial.}	
	\includegraphics[scale=3]{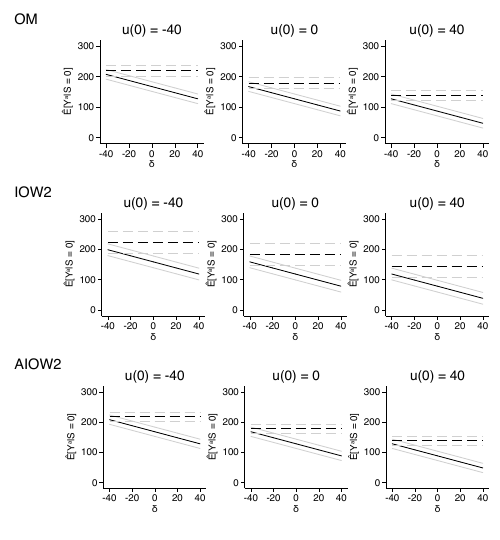}
	\label{fig:sens_anal_HALT_C}
	\caption*{AIOW2 = augmented inverse odds weighting estimator with normalized weights; IOW2 = inverse odds weighting with normalized weights; OM = outcome model-based estimator. Results are shown as point estimates (black lines) and corresponding 95\% confidence intervals (gray lines) for potential outcome means under treatment $a=1$ (solid lines) and $a=0$ (dashed lines). Results for IOW1 and AIOW1 were similar to IOW2 and AIOW2, respectively, and are shown in the Appendix.}
\end{figure}

\clearpage
\linespread{1.25}
\appendix

\renewcommand{\theequation}{A.\arabic{equation}}
\setcounter{equation}{0}

\section{Identification, estimation, and sensitivity analysis for potential outcome means in the entire target population}\label{appendix:all_eligible}

\vspace{0.2in}

We now summarize methods for the identification, estimation, and sensitivity analysis for the potential outcome means in the entire target population, $\E[Y^a]$. These methods apply to nested trial designs (i.e., when the randomized trial is nested in a cohort sampled from the target population).

\subsection*{Identifiability conditions}

\noindent
\emph{(I) Consistency:}  If $ A_i = a$, then $ Y_i = Y^{a}_i.$ Again, implicit in our notation is the assumption that the invitation to participate in the trial and trial participation itself do not affect the outcome except through treatment assignment. 

\noindent
\emph{(II) Mean exchangeability in the trial (over $A$):} $ \E [ Y^{a} | X, S = 1, A = a] = \E [ Y^{a} | X , S = 1 ].$

\noindent
\emph{(III) Positivity of treatment assignment:} $ \Pr[A=a | X = x, S=1] > 0$ for every  $a$ and every $x$ with positive density among randomized individuals, $f_{X|S}(x| S = 1) > 0.$
 
\noindent
\emph{(IV) Mean generalizability (exchangeability over S):} $ \E [ Y^{a} | X, S = 1 ] = \E [ Y^{a} | X].$
 
\noindent
\emph{(V) Positivity of trial participation:} $\Pr[S = 1 | X = x] >0$ for every $x$ with positive density in the population, $f_X(x) > 0.$

\subsection*{Identification}

Under the identifiability conditions listed above, the potential outcome mean among the target population is identifiable as
\begin{align}
	\E [Y^a ] &= \E\! \big[\! \E [Y  | X , S = 1 , A = a ] \big] \label{eq:identitification_ALL}  \\
		&= \E \left[ \dfrac{S I (A = a) Y }{\Pr[S = 1| X] \Pr[A = a | X, S = 1 ]} \right] \label{eq:identitification_ALL_ipw},
\end{align}
where the second equality only requires the positivity conditions \cite{dahabreh2018transporting}.

\subsection*{Estimation}

\noindent
\textbf{Potential outcome means:} The identification results above suggest several ways for estimating the potential outcome mean in the target population. In this section, we use the same notation as in the main text, except that we use the tilde symbol to denote estimators. Additional information about the behavior of estimators when the mean transportability assumption holds are available in \cite{dahabreh2018generalizing}.

\vspace{0.1in}

\noindent
\emph{g-formula estimator:} We can use the sample analog estimator based on (\ref{eq:identitification_ALL}),
\begin{equation}\label{estimator:standard}
	\widetilde \mu_{\text{\scalebox{\sizevar}{OR}}}(a) = \dfrac{1}{n} \sum\limits_{i=1}^{n} \widehat g_a(X_i) .
\end{equation}

\noindent
\emph{Inverse probability (IP) weighting:} We can use an estimator based on (\ref{eq:identitification_ALL_ipw})
\begin{equation} \label{estimator:ipw1}
	\widetilde \mu_{\text{\scalebox{\sizevar}{IPW1}}}(a) = \dfrac{1}{n} \sum\limits_{i=1}^{n} \widetilde w_a(X_i, S_i, A_i) Y_i ,
\end{equation}
with $$ \widetilde w_a(X_i, S_i, A_i) = \dfrac{S_i I(A_i = a)}{\widehat p(X_i) \widehat e_a(X_i)}.$$
We can normalize the weights to sum to 1,
\begin{equation} \label{estimator:ipw2}
	\widetilde \mu_{\text{\scalebox{\sizevar}{IPW2}}}(a) = \left\{ \sum\limits_{i=1}^{n} \widetilde w_a(X_i, S_i, A_i) \right\}^{-1} \sum\limits_{i=1}^{n} \widetilde w_a(X_i, S_i, A_i) Y_i.
\end{equation}

\noindent
\emph{Augmented IP weighting:} To improve the efficiency of the IP weighting estimator and allow for robustness to misspecification of the outcome mean or probability of trial participation models, we can use an augmented (doubly robust) IP weighting estimator \cite{dahabreh2018generalizing},
\begin{equation} \label{estimator:aipw1}
\begin{split}
\widetilde{\mu}_{\text{\scalebox{\sizevar}{AIPW1}}} (a) =  \frac{1}{n} \sum_{i=1}^{n} \Big\{ \widetilde w_a(X_i, S_i, A_i) \big\{ Y_{i} - \widehat g_a (X_i) \big\} + \widehat g_a (X_i)  \Big\},
\end{split}
\end{equation}
Again, it is often better to normalize the weights to sum to 1, 
\begin{equation} \label{estimator:aipw2}
\begin{split}
\widetilde{\mu}_{\text{\scalebox{\sizevar}{AIPW2}}}(a) &=  \left\{ \sum_{i = 1}^{n}\widetilde w_a(X_i, S_i, A_i) \right\}^{-1} \sum_{i=1}^{n}   \widetilde w_a(X_i, S_i, A_i) \big\{ Y_{i} - \widehat g_a (X_i) \big\} + \dfrac{1}{n} \sum\limits_{i=1}^{n} \widehat g_a(X_i).
\end{split}
\end{equation}

\vspace{0.1in}
\noindent
\textbf{Treatment effects:} We can use the above potential outcome mean estimators, to estimate the average treatment effect in the target population, $\E [Y^1 - Y^0]$, by differencing.

\subsection*{Inference} As usual, confidence intervals for the estimated potential outcome means and mean differences can be obtained by the usual ``sandwich'' approach \cite{Stefanski2002} or by bootstrap methods \cite{efron1994introduction}.

\subsection*{Sensitivity analysis} 

As in the main text, we parameterize violations of the transportability assumption using the bias functions \cite{robins2000c,brumback2004sensitivity,robins1999association}
\begin{equation*}
	u(a, X ) 	= \E[Y^a | X , S = 1 ] - \E[Y^a | X , S = 0 ].
\end{equation*}

The parameterization allows us to re-express the conditional potential outcome mean under treatment $a$,
\begin{equation*}
	\begin{split}
	\E  [Y^a | X ]  	&= \E [Y^a | X , S = 1] \Pr[S = 1 | X] + \E [Y^a | X , S = 0] \Pr[S = 0 |  X]  \\
							&= \E [Y | X,  S = 1, A = a]  \Pr[S = 1 | X]  \\
									& \quad \quad + \big\{    \E[Y | X , S = 1 , A = a]    -   u(a, X )  \big\} \Pr[S = 0 |  X]   \\
							&= \E [Y | X, S = 1, A = a] -   u(a, X ) \Pr[S = 0 |  X],
	\end{split}
\end{equation*}
where the first equality follows from the law of total expectation; the second from consistency and mean exchangeability in the trial together with the definition of $u(a, X)$; and the third from the fact that $S$ is binary.

Taking expectations over the distribution of $X$ in the target population, 
\begin{equation}\label{eq:basis_for_sens}
	\begin{split}
	\E [Y^a] &= \E\! \big[\! \E [Y | X , S = 1, A = a] -   u(a, X ) \Pr[S = 0 |  X] \big] \\
			&= \E\! \big[\! \E [Y | X , S = 1, A = a] \big] - \E\! \big[ u(a, X ) \Pr[S = 0 |  X] \big] \\
			&= \E\! \big[\! \E [Y | X , S = 1, A = a] \big] - \E\! \big[ (1 - S) u(a, X )  \big],
	\end{split}
\end{equation}
where the last equality follows from the law of total expectation.

\vspace{0.1in}
\noindent
\textbf{Sensitivity analysis estimators:} The result in (\ref{eq:basis_for_sens}) suggests simple approaches for sensitivity analysis using the potential outcome mean estimators in the previous section.

\vspace{0.1in}
\noindent
\emph{g-formula estimator:} We can incorporate the bias correction function directly,
\begin{equation}\label{estimator:standard_bc}
	\widetilde \mu_{\text{\scalebox{\sizevar}{OR}}}^{\text{\scalebox{\sizevar}{bc}}}(a) = \dfrac{1}{n} \sum\limits_{i=1}^{n} \Big\{ \widehat g_a(X_i) - (1 - S_i) u(a, X_i) \Big\}.
\end{equation}

\vspace{0.1in}
\noindent
\emph{IP weighting:} Similarly, we can modify the IP weighting estimators as
\begin{equation} \label{estimator:ipw1_bc}
	\widetilde \mu_{\text{\scalebox{\sizevar}{IPW1}}}^{\text{\scalebox{\sizevar}{bc}}}(a) = \dfrac{1}{n} \sum\limits_{i=1}^{n} \Bigg\{\widetilde w_a(X_i, S_i, A_i) Y_i - (1 - S_i)u(a, X_i)  \Bigg\},
\end{equation}
or, normalizing the weights to sum to 1, 
\begin{equation} \label{estimator:ipw2_bc}
	\widetilde \mu_{\text{\scalebox{\sizevar}{IPW2}}}^{\text{\scalebox{\sizevar}{bc}}}(a) = \left\{ \sum\limits_{i=1}^{n} \widetilde w_a(X_i, S_i, A_i) \right\}^{-1} \sum\limits_{i=1}^{n}\widetilde w_a(X_i, S_i, A_i) Y_i-  \dfrac{1}{n} \sum\limits_{i=1}^{n} (1 - S_i) u(a, X_i)  .
\end{equation}

In the next section of this Appendix, we show how the bias correction for the IP weighting estimators can be easily implemented using standard regression software, by simply re-coding the outcome values to directly incorporate the bias correction function $u(a, X)$.

\vspace{0.1in}
\noindent
\emph{Augmented IP weighting:} We can incorporate the bias correction functions in the augmented IP weighting estimators,
\begin{equation} \label{estimator:aipw1_bc}
\begin{split}
\widetilde{\mu}_{\text{\scalebox{\sizevar}{AIPW1}}}^{\text{\scalebox{\sizevar}{bc}}} (a) =  \frac{1}{n} \sum_{i=1}^{n} \Big\{ \widetilde w_a(X_i, S_i, A_i) \big\{ Y_{i} - \widehat g_a (X_i) \big\} + \widehat g_a (X_i)  - (1 - S_i) u(a, X_i) \Big\},
\end{split}
\end{equation}
or, normalizing the weights to sum to 1,
\begin{equation} \label{estimator:aipw2_bc}
\begin{split}
\widetilde{\mu}_{\text{\scalebox{\sizevar}{AIPW2}}}^{\text{\scalebox{\sizevar}{bc}}}(a) &=  \left\{ \sum_{i = 1}^{n} \widetilde w_a(X_i, S_i, A_i) \right\}^{-1} \sum_{i=1}^{n} \widetilde w_a(X_i, S_i, A_i) \big\{ Y_{i} - \widehat g_a (X_i) \big\} \\
  &\quad\quad\quad\quad\quad\quad + \dfrac{1}{n} \sum\limits_{i=1}^{n} \big\{ \widehat g_a(X_i) - (1 - S_i) u(a, X_i) \big\}.
\end{split}
\end{equation}

These augmented IP weighting estimators retain the double robustness property of their non-bias corrected counterparts, in the sense that, when the bias correction function is correctly specified, they produce valid results when either the model of the outcome mean among randomized individuals or for the probability of trial participation is correctly specified.

\vspace{0.1in}
\noindent
\textbf{Sensitivity analysis for treatment effects:} As when the transportability assumption holds, sensitivity analysis for treatment effects can be obtained by differencing the appropriate bias-corrected estimators.

\subsection*{Inference for sensitivity analysis} 

The general theory for sensitivity analysis for semi-parametric models in \cite{robins2000c} shows that confidence intervals for the estimated bias-corrected potential outcome means and mean differences can also be obtained by ``sandwich'' or bootstrap \cite{efron1994introduction} methods, while accounting for the uncertainty in estimating the parameters of the model for the probability of trial participation.

\clearpage
\section{Sensitivity analysis with bias-corrected \\ outcomes} 

We now demonstrate the connection between our approach for sensitivity analysis and the method of using ``bias corrected outcomes'' in marginal structural models \cite{robins1999association, brumback2004sensitivity}.

\vspace{0.1in}
\noindent
\textbf{Sensitivity analysis with bias-corrected outcomes for $\E[Y^a | S = 0]$:} Define the bias corrected outcomes $$ Y_i^* =  Y_i - u(A_i, X_i) $$ and assume that bias functions $u(a, X)$ have been properly chosen, so that, for every $a$, $$ u(a, X) = \E[Y^a | X , S = 1 ] - \E[ Y^a | X , S = 0].$$ The estimator
\begin{equation*}
	\widehat \mu_{\text{\scalebox{\sizevar}{IOW}}}^{\text{\scalebox{\sizevar}{bc obs}}} (a) = \left\{ \sum\limits_{i=1}^{n} \widehat w_a(X_i, S_i, A_i) \right\}^{-1} \sum\limits_{i=1}^{n} \widehat w_a(X_i, S_i, A_i) Y_i^*,
\end{equation*}
with $$\widehat w_a(X_i, S_i, A_i) = S_i I(A_i = a) \dfrac{1 - \widehat p(X_i)  }{\widehat p(X_i) \widehat e_a(X_i)},$$ converges in probability to the same limit as the estimators in (\ref{estimator:iow1_bc}) and (\ref{estimator:iow2_bc}).
The bias-corrected outcome estimator above can be obtained by running a weighted least squares regression of $Y^*$ on $A$ with weights $\widehat w_a(X_i, S_i, A_i)$.

\vspace{0.1in}
\noindent
\textbf{Sensitivity analysis with bias-corrected outcomes for $\E[Y^a]$:} For each trial participant, define the \emph{bias-corrected outcome} $$ \widetilde Y_i^* = Y_i - u(A_i, X_i) \big\{ 1 -  \widehat p(X_i) \big\}.$$ The estimator 
\begin{equation*}\label{mean_bias_cor}
	\widehat \mu_{\text{\scalebox{\sizevar}{IPW}}}^{\text{\scalebox{\sizevar}{bc obs}}}(a) = \left\{ \sum\limits_{i=1}^{n} \widetilde w_a(X_i, S_i, A_i) \right\}^{-1} \sum\limits_{i=1}^{n} \widetilde w_a(X_i, S_i, A_i)  \widetilde Y_i^*,
\end{equation*}
with $$\widetilde w_a(X_i, S_i, A_i) =  \dfrac{S_i I(A_i = a)}{\widehat p(X_i) \widehat e_a(X_i)},$$ converges in probability to the same limit as estimators (\ref{estimator:ipw1_bc}) and (\ref{estimator:ipw2_bc}). The bias-corrected outcome estimator above can be obtained by running a weighted least squares regression of $\widetilde Y^*$ on $A$ with weights $\widetilde w_a(X_i, S_i, A_i).$

\vspace{0.1in}
\noindent
\textbf{Inference for sensitivity analysis using bias corrected outcomes:} Inference for the potential outcome means and treatment effects when using bias-corrected observations is best obtained using bootstrap methods \cite{brumback2004sensitivity}.

\clearpage
\section{Additional HALT-C results}
\setcounter{table}{0} 
\renewcommand\thetable{C.\arabic{table}} 
\setcounter{figure}{0}
\renewcommand{\thefigure}{C.\arabic{figure}}

In this Appendix we report additional results from the re-analysis of data from the HALT-C trial.

\begin{landscape}
\begin{table}[p]
  \centering
{\small

	\caption{Patient characteristics in HALT-C transportability analyses, by center ($S$) and treatment assignment ($A$). Results are reported as a proportion for binary variables or as mean (standard deviation) for continuous variables.}	
\begin{tabular}{lc|cc|c|cc}
\toprule
Variable                                                                             & \begin{tabular}[c]{@{}c@{}}S=1 \\ (n = 94)\end{tabular} & \begin{tabular}[c]{@{}c@{}}S=1, A=1 \\ (n = 46)\end{tabular} & \begin{tabular}[c]{@{}c@{}}S=1, A=0\\ (n = 48)\end{tabular} & \begin{tabular}[c]{@{}c@{}}S=0\\ (n = 92)\end{tabular} & \begin{tabular}[c]{@{}c@{}}S=0, A=1\\ (n = 46)\end{tabular} & \begin{tabular}[c]{@{}c@{}}S=0, A=0\\ (n = 46)\end{tabular} \\ \midrule
Baseline platelets                                                                   & 169.5 (69.9)                                            & 182.3 (72.1)                                               & 157.1 (66.0)                                                & 182.9 (64.2)                                           & 168.7 (58.2)                                                & 197.1 (67.3)                                                \\
Age                                                                                  & 50.3 (6.4)                                              & 50.2 (6.3)                                                   & 50.4 (6.5)                                                  & 49.1 (56.0)                                            & 50.1 (5.4)                                                  & 48.0 (6.3)                                                  \\
Female                                                                               & 0.22                                                    & 0.24                                                         & 0.21                                                        & 0.23                                                   & 0.13                                                        & 0.33                                                        \\
\begin{tabular}[c]{@{}l@{}}History of pegylated \\ interferon treatment\end{tabular} & 0.31                                                    & 0.24                                                         & 0.38                                                        & 0.16                                                   & 0.15                                                        & 0.17                                                        \\
White-Non-Hispanic                                                                   & 0.67                                                    & 0.57                                                         & 0.77                                                        & 0.74                                                   & 0.74                                                        & 0.74                                                        \\
Baseline WBC                                                                         & 6.0 (1.9)                                               & 6.3 (2.1)                                                    & 5.7 (1.6)                                                   & 5.9 (1.7)                                              & 6.0 (1.8)                                                   & 5.8 (1.7)                                                   \\
\begin{tabular}[c]{@{}l@{}}Used needles or \\ recreational drugs\end{tabular}        & 0.57                                                    & 0.59                                                         & 0.56                                                        & 0.49                                                   & 0.52                                                        & 0.46                                                        \\
Transfusion history                                                                  & 0.34                                                    & 0.37                                                         & 0.31                                                        & 0.39                                                   & 0.41                                                        & 0.37                                                        \\
BMI                                                                                  & 30.3 (5.8)                                              & 30.5 (6.7)                                                   & 30.1 (4.9)                                                  & 29.8 (5.5)                                             & 29.7 (4.7)                                                  & 29.9 (6.3)                                                  \\
Baseline creatinine                                                                  & 0.86 (0.16)                                             & 0.87 (0.18)                                                  & 0.85 (0.141)                                                & 0.83 (0.15)                                            & 0.85 (0.14)                                                 & 0.80 (0.16)                                                 \\
Smoking (never)                                                                      & 0.22                                                    & 0.26                                                         & 0.19                                                        & 0.27                                                   & 0.24                                                        & 0.30                                                       
 \\     \bottomrule                                               
\end{tabular}
	\label{tab:descriptives}
  }
\end{table}

\end{landscape}

\clearpage
\begin{table}[p]
  \centering
  \caption{``Base case'' transportability analyses using center $S=0$ as the ``target center'' and assuming conditional mean transportability holds ($u(0)=0 $ and $\delta = 0$); 95\% confidence intervals are given in parentheses.} 
\begin{tabular}{llll}
\toprule
\textbf{Estimator} & $a=1$            & $a=0$            & \textbf{Mean difference} \\ \midrule
OM                 & 126.81 (111.03, 142.59) & 178.38 (161.49, 195.27) & -51.57 (-70.49, -32.65)  \\ 
IOW1               & 122.60 (99.33, 145.87)  & 175.66 (120.76, 230.56) & -53.06 (-116.47, 10.35)  \\ 
IOW2               & 118.26 (98.91, 137.61)  & 182.87 (147.63, 218.11) & -64.61 (-100.55, -28.67) \\ 
AIOW1              & 127.55 (112.13, 142.97) & 177.27 (161.87, 192.67) & -49.72 (-67.44, -32.00)  \\ 
AIOW2              & 127.53 (112.11, 142.95) & 177.23 (161.87, 192.59) & -49.70 (-67.37, -32.03)  \\ \bottomrule
\end{tabular}
  \label{tab:treatment_effects_trans}
\end{table}

\clearpage
\begin{table}[p]
	\centering
	\caption{Crude and adjusted analyses in each research center contributing data to the transportability analyses; 95\% confidence intervals are given in parentheses.}	
\begin{tabular}{lllll}
\toprule
\textbf{Center}        & \textbf{Estimate} & \textbf{$a=1$}          & \textbf{$a=0$}          & \textbf{Mean difference} \\ \midrule
\multirow{2}{*}{S = 1} & Crude             & 130.87 (110.6, 151.13)  & 155.77 (135.93, 175.61) & -24.9 (-53.26, 3.46)     \\ 
                       & Adjusted          & 130.47 (112.25, 148.69) & 166.86 (150.2, 183.52)  & -36.39 (-55.64, -17.14)  \\ \hline
\multirow{2}{*}{S = 0} & Crude             & 119.87 (97.54, 142.2)   & 204.8 (182.48, 227.13)  & -84.93 (-116.51, -53.36) \\ 
                       & Adjusted          & 128.37 (115.62, 141.12) & 190.47 (168.07, 212.87) & -62.1 (-85.19, -39.01)   \\ \bottomrule
\end{tabular}
	\label{tab:treatment_effects_no_trans}
\end{table}

\clearpage
\begin{figure}[p]
  \centering
  \caption{Sensitivity analysis using the inverse odds of participation estimator with non-normalized weights (IOW1) and the augmented inverse odds of participation estimator with non-normalized weights (AIOW1).}  
  \includegraphics[scale=3]{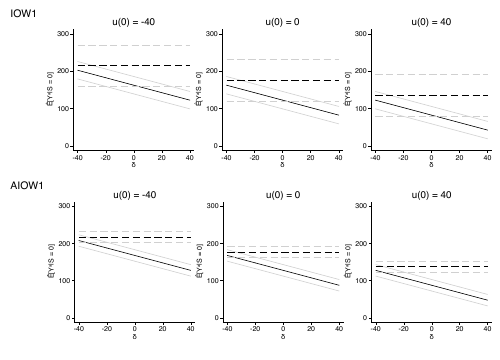}
  \label{fig:appendix_A1}
\end{figure}

\clearpage
\section{Sensitivity analysis with covariate-dependent bias functions}\label{appendix:covariate_dependent_bias_functions}
\setcounter{figure}{0}
\renewcommand{\thefigure}{D.\arabic{figure}}

To illustrate the application of the sensitivity analysis methods when the bias functions depend on covariates, we considered a sensitivity analysis where the bias functions depended on sex (coded as male vs. female). Without loss of generality, let the first element of $X$, $X_1$, be a random variable that takes the value 1 for male and 0 for female individuals. We set up the bias function for $a=0$ as
\begin{equation*}
	\begin{split}
u(0, X) &\equiv  I(X_1 = 1) \times u(0)  +  0.8 \times I(X_1 = 0) \times u(0),
	\end{split}
\end{equation*}
such that the magnitude of assumption violations is smaller by 20\% among female individuals versus male individuals. Similarly, we set the difference of the bias functions to
\begin{equation*}
	\begin{split}
\delta(X) &\equiv  I(X_1 = 1) \times \delta  +  0.8 \times I(X_1 = 0) \times \delta.
	\end{split}
\end{equation*}

\begin{figure}[ht!]
	\centering
	\caption{Sensitivity analysis results for transporting inferences between two research centers participating in the HALT-C trial using bias functions that depend on baseline covariates.}	
	\includegraphics[scale=3]{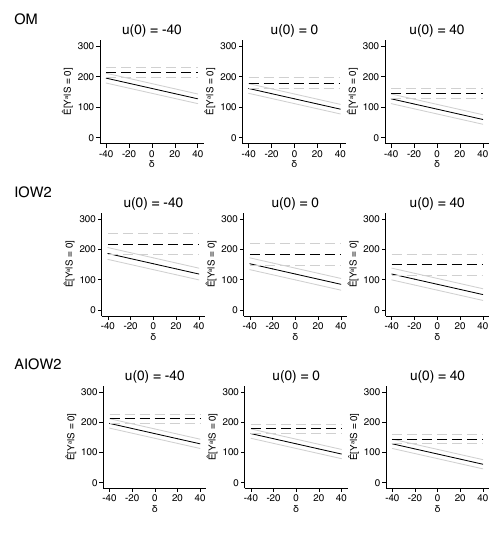}
	\label{fig:sens_anal_HALT_C_cov_dep}
	\caption*{AIOW2 = augmented inverse odds weighting estimator with normalized weights; IOW2 = inverse odds weighting with normalized weights; OM = outcome model-based estimator. Results are shown as point estimates (black lines) and corresponding 95\% confidence intervals (gray lines) for potential outcome means under treatment $a=1$ (solid lines) and $a=0$ (dashed lines). Results for IOW1 and AIOW1 were similar to IOW2 and AIOW2, respectively, and are not shown here.}
\end{figure}

Figure \ref{fig:sens_anal_HALT_C_cov_dep} presents sensitivity analysis results using the bias functions described above. Comparing the results plotted in Figure \ref{fig:sens_anal_HALT_C_cov_dep} against those in Figure \ref{fig:sens_anal_HALT_C}, we see that sensitivity is slightly reduced. The reason is that violations of transportability assumption in Figure \ref{fig:sens_anal_HALT_C_cov_dep} are assumed to be the same among male but smaller (by 20\%) among female individuals, compared to the corresponding values used for Figure \ref{fig:sens_anal_HALT_C}.

\clearpage

\end{document}